\documentclass[preprint2]{aastex}
\usepackage{epsfig}
\usepackage{natbib}
\usepackage{amsmath}

\newcommand{\kms}{km~s$^{-1}$}
\newcommand{\Msun}{M_{\odot}}

\begin{document}
\title{The Arrowhead Mini-Supercluster of Galaxies}

\author{Daniel Pomar\`ede}
\affil{Institut de Recherche sur les Lois Fondamentales de l'Univers, CEA/Saclay}
\and
\author{R. Brent Tully,}
\affil{Institute for Astronomy, University of Hawaii, 2680 Woodlawn Drive,
 Honolulu, HI 96822, USA}
\and
\author{Yehuda Hoffman}
\affil{Racah Institute of Physics, Hebrew University, Jerusalem 91904, Israel}
\and
\author{H\'el\`ene M. Courtois}
\affil{Universit\'e Claude Bernard Lyon I, Institut de Physique Nucl\'eaire, Lyon, France}

\begin{abstract}
Superclusters of galaxies can be defined kinematically from local evaluations of the velocity shear tensor.  The location where the smallest eigenvalue of the shear is positive and maximal defines the center of a basin of attraction.  Velocity and density fields are reconstructed with Wiener Filter techniques.  Local velocities due to the density field in a restricted region can be separated from external tidal flows, permitting the identification of boundaries separating inward flows toward a basin of attraction and outward flows.
This methodology was used to define the Laniakea Supercluster that includes the Milky Way.  Large adjacent structures include Perseus$-$Pisces, Coma, Hercules, and Shapley but current kinematic data are insufficient to capture their full domains.   However there is a small region trapped between Laniakea, Perseus$-$Pisces, and Coma that is close enough to be reliably characterized and that satisfies the kinematic definition of a supercluster.  Because of its shape, it is given the name the Arrowhead Supercluster.  This entity does not contain any major clusters.  A characteristic dimension is $\sim 25$~Mpc  and the contained mass is only $\sim 10^{15}~\Msun$.

\bigskip
\end{abstract}

\smallskip
\section{Introduction}

Velocity field information can be used to give a quantitative delineation of superclusters of galaxies.  Large scale structure in the cosmic web is interconnected by a network of filaments but shear in flow patterns inform efforts to isolate adjacent distinct basins of attraction.   

The radial component of the peculiar velocities of a sampling of available galaxies is averaged and interpreted in terms of a continuous three-dimensional velocity field using the Wiener Filter with constrained realizations \citep{2012ApJ...744...43C}.  The procedure involves a Bayesian fit to the assumed conditions of a Gaussian initial distribution of fluctuations that grow linearly and obey a specified power spectrum.  Each constrained realization conforms to the requirements imposed by the observations and the power spectrum.  A suite of constrained simulations gives similar velocity and density fields where the data are dense and accurate but gives poorly correlated fields where the data are sparse and inaccurate.  A mean of many constrained realizations provides renditions of the velocity and density field that are detailed nearby but degrade in the average to the mean density beyond the data zone.

Knowledge of the density field allows for a separation into locally induced velocity fields and larger scale tidally induced flows.  The region under consideration as local can be specified and densities external to the region are set to the mean.  The density within this region induces velocity patterns that, in linear theory, is the Poisson relationship between densities and the divergence of the velocity field.  The vector subtraction of this local velocity field from the global velocity field gives the tidal component.

Knowledge of the three-dimensional velocity field allows the computation of the velocity shear at any location.  The ordered eigenvalues of the shear tensor identifies whether the location can be characterized as a knot, filament, sheet, or void depending on the collapse or expansion properties in the three orthogonal eigenvector directions.  The V-web is created from surfaces bounding knots, filaments, and sheets.  Non-overlapping basins of gravitational attraction can be identified from velocity flow patterns.  Locations of divergence in flows toward neighboring basins delineate structural boundaries, metaphorically similar to the separation between terrestrial watersheds.  This characteristic provides a useful quantitative way to define galaxy superclusters. 

Recently, a Wiener Filter analysis of the motions of galaxies derived from the Cosmicflows-2 (CF2) compendium of galaxy distances \citep{2013AJ....146...86T} led to the identification of the boundaries of the supercluster that we live in.  This structure was given the name Laniakea Supercluster \citep{2014Natur.513...71T}.    

The major structures bordering Laniakea are Perseus-Pisces \citep{1988lsmu.book...31H}, Coma \citep{1986ApJ...302L...1D}, Hercules \citep{1979ApJ...234..793T} and Shapley \citep{1989Natur.342..251R}.   In addition, there turns out to be a curious minor structure at the Laniakea boundary that satisfies the condition of a local basin of attraction.  This feature is given attention in the present study.  It is given the name the "Arrowhead" Supercluster.

\begin{figure}[htbp]
\begin{center}
\includegraphics[scale=1]{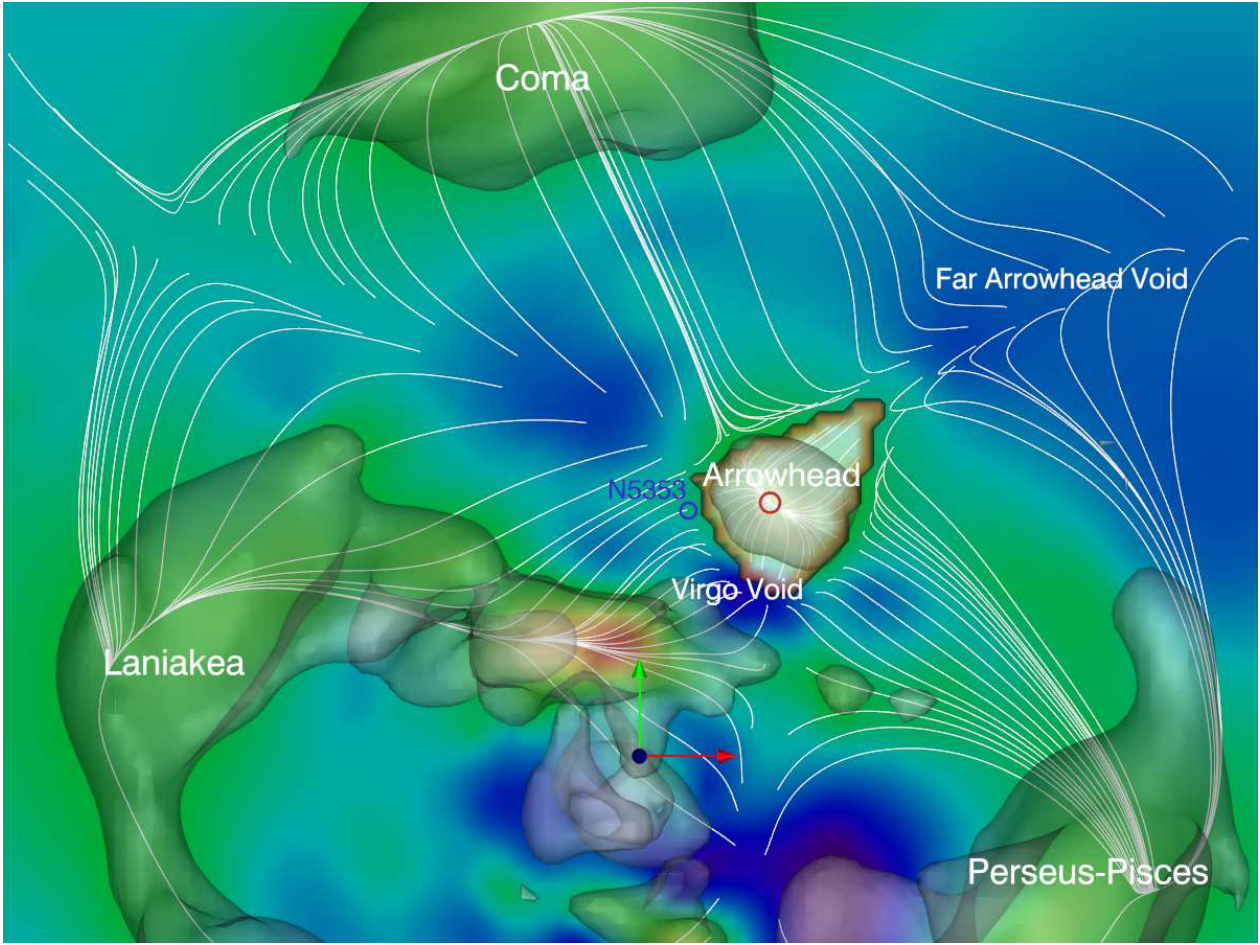}
\caption{Arrowhead supercluster seen in dark outline in a slice in supergalactic coordinates: horizontal $-6470 < SGX < +6280$~\kms; vertical $-1900 < SGY < +7660$~\kms; thickness $-500 < SGZ < +1500$~\kms.  The Milky Way galaxy is at the origin of the red and green arrows, 1000 \kms\ in length, red toward +SGX and green toward +SGY.  A blue arrow toward +SGZ, directed out of the page here, is seen in subsequent figures.  Colored contours represent density levels as reconstructed by the Wiener Filter analysis with red at high densities, green intermediate, and blue at low densities.  The semi-transparent surfaces represent an intermediate contour in density.  The white lines illustrate velocity flow patterns initiated at selected seed positions showing flows both internal to Arrowhead and away, alternatively toward Laniakea, Perseus-Pisces, or Coma.  The Arrowhead Supercluster is bracketed by the small Virgo Void and extensive Far Arrowhead Void in this projection.
This scene appears at 00:40 in the accompanying video.
}
\label{sgxy}
\end{center}
\end{figure}

The Arrowhead is balanced between three large gravitational competitors: Laniakea, Perseus-Pisces, and Coma.  A view is seen in Figure~\ref{sgxy}.  Flows around the edges are directed toward the three large structures but local flows are confined within the restricted region of the dark outline.

The center of the Arrowhead Supercluster is at roughly $12^h05^m+58^{\circ}$; $\ell=+134^{\circ},b=+57^{\circ}$; $SGL=+57^{\circ},SGB=+7^{\circ}$; distance $\sim 2700$~\kms.  The near side of the region is included in the {\it Nearby Galaxies Atlas} \citep{1987nga..book.....T} as structures called the Canes Venatici$-$Camelopardalis Cloud and the Bo\"otes Cloud.  The limit of that early atlas is 3,000~\kms\ so includes only the front side of the Arrowhead Supercluster.  The most prominent individual group is at the periphery of the supercluster: the NGC~5353/54 Group with a mass $3\times10^{13}~\Msun$ \citep{2008AJ....135.1488T} at a revised distance of 35 Mpc.  The region being studied is at high galactic latitude so there are no serious issues with incompleteness due to obscuration.

As an overview it is to be noted that
prominent filaments run from the vicinity of the NGC~5353/54 Group toward both the Coma and Perseus$-$Pisces structures and a minor filament runs toward the Virgo Cluster in Laniakea Supercluster.  The minor filament in the foreground of the Arrowhead running toward the Virgo Cluster is called the Canes Venatici Spur in the {\it Nearby Galaxies Atlas}.  The important Canes Venatici$-$Camelopardalis (CVn$-$Cam) Cloud starts near the NGC~5353/54 Group within the Arrowhead Supercluster and heads toward the Perseus$-$Pisces structure.  As will be described further along, at a certain point within CVn$-$Cam the flow within the filament becomes dominated by Perseus$-$Pisces.  Finally, as demonstrated in Figure~2 in \citet{2008AJ....135.1488T}, there is a prominent filament running from the vicinity of the NGC~5353/54 Group to the Coma Cluster.  The flow along the length of this filament is toward Coma.  These features are seen in plots by \citet{2013AJ....146...69C}.  The CVn$-$Cam Cloud is seen in Fig.~7 of that paper slanting down toward the right toward Perseus$-$Pisces.  The filament that has been described connecting to the Coma Cluster is seen in Fig.~6.

\onecolumn
\twocolumn

This article is accompanied by a video.\footnote{http://irfu.cea.fr/arrowhead}  If an image can be worth a thousand words, the video is worth many thousand.

\section{Distances}

The velocity and density fields, already shown as a tease in Fig.~\ref{sgxy}, are derivatives of distance measurements.  The cosmic expansion velocity expected at a measured distance can be subtracted from the observed velocity.  The residual is the peculiar velocity, the radial component of the motion of the target galaxy, supposed to be in response to discrete components of the matter distribution.  The current analysis makes use of the {\it Cosmicflows-2} (CF2) catalog of distances to 8190 galaxies  \citep{2013AJ....146...86T}.  The putative Arrowhead structure lies at high galactic latitude ($b>40^{\circ}$) in a well studied part of the northern sky.  In a generous region encompassing the Arrowhead ($9^h < RA < 16.5^h$, $25^{\circ}<Dec<80^{\circ}$) there are 871 distance estimates in CF2 within $cz=10,000$~\kms.  
In the immediate vicinity and within the Arrowhead there are 126 distance measurements, 81 of these in 22 groups where 2 to 10 estimates can be averaged.  Distance uncertainties range from 20\% rms for most of the 45 singles, to $6-10\%$ for 10 well populated groups, with intermediate uncertainties for another dozen pairs or triples.

\begin{figure}[t]
\begin{center}
\includegraphics[scale=0.4]{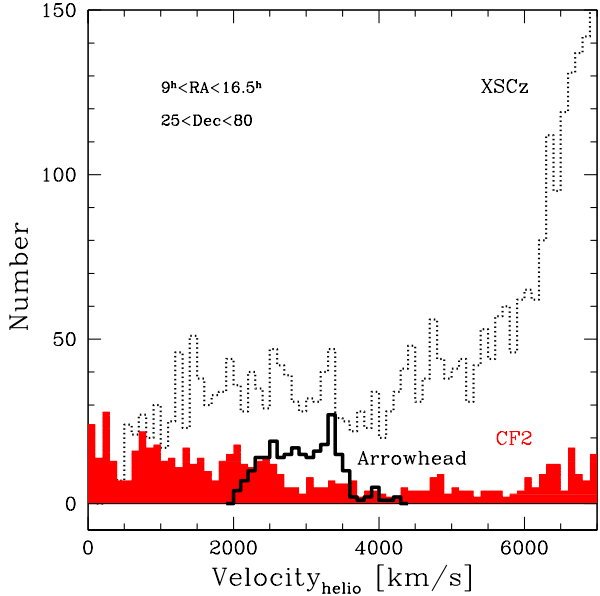}
\caption{Filled red histogram: number of galaxies in 100 \kms\ bins with CF2 distance measurements in a region enclosing the Arrowhead Supercluster.  Dotted histogram: number of galaxies in the same region from XSCz. Heavy open histogram: XSCz galaxies within Arrowhead Supercluster.
}
\label{histv}
\end{center}
\end{figure}

Details of the content of this section of sky are provided graphically in Figure~\ref{histv}.  The filled histogram shows the distribution of velocities of galaxies with measured distances.  The dotted outer histogram is built from those galaxies in the same part of the sky in the 2MASS Extended Source Catalog \citep{2000AJ....119.2498J} with known redshifts (XSCz).  The 243 galaxies in the XSCz that lie within the Arrowhead contribute to the heavy solid histogram.

The details regarding the CF2 distance measurements are provided in \citet{2013AJ....146...86T} and the distances are available at the Extragalactic Distance Database \citep{2009AJ....138..323T}.\footnote{http://edd.ifa.hawaii.edu}  A few comments specific to the region of the sky are in order.  As already noted, the area of interest is in the northern sky, well separated from the zone of obscuration.  From the redshift distribution of the XSCz sample it can be seen that there is a severe deficiency of galaxies in the interval $3500<V_{helio}<6000$~\kms.  The dramatic rise at 6000~\kms\ is due to the onset of the Great Wall in the vicinity of the Coma Cluster.  There are plenty of galaxies to the foreground of the Arrowhead feature because the line-of-sight is through busy parts of the historic Local Supercluster in Canes Venatici and Ursa Major.

There is a continuous representation of galaxies with distance measurements across the domain of interest.  The drop in number beyond 3000~\kms\ is largely due to the onset of the background void.  One of the 7 Tully-Fisher samples in the CF2 catalog has an imposed limit of 3300~\kms\ but the other components of CF2, including 5 other methodologies, are blind to this limit and the Wiener Filter analysis is not expected to be strongly affected.  The Wiener Filter reconstruction discussed in the next section is robustly constrained within 6,000~\kms, encompassing the volume of present interest.

\section{The Wiener Filter Analysis}

The Wiener Filter methodology for the reconstruction of velocity and density fields from peculiar velocity information has been discussed in a sequence of papers  \citep{1995ApJ...449..446Z, 1999ApJ...520..413Z, 2012ApJ...744...43C}.  The specific reconstruction based on CF2 distances was described in an article about the Laniakea Supercluster \citep{2014Natur.513...71T}.  The focus here is on a relatively small region and it is useful to separate the velocity field into local patterns in the domain of interest and the residual tidal component  \citep{2001astro.ph..2190H}.

The structure that is being given attention was already seen as curious with the publication of the Nearby Galaxies Atlas \citep{1987nga..book.....T}.  The backbone of the CVn-Cam Cloud slants at an angle with respect to adjacent features that lie parallel to the supergalactic equator and instead of being directed toward the nearby Virgo Cluster as might be anticipated, it runs normal to the Virgo direction.  Now from the Wiener Filter density and velocity field reconstruction, it was found that the Laniakea entity extends up to, but evidently not including, the CVn-Cam filament.  The CVn-Cam filament runs along the edge of the Arrowhead region on the side toward our viewing position.  There are flows at the ends of this filament toward both Perseus-Pisces and Coma and a spur off the side toward Laniakea.  This nearby but apparently dynamically autonomous region deserved a closer look.

\begin{figure}[!]
\begin{center}
\includegraphics[scale=0.82]{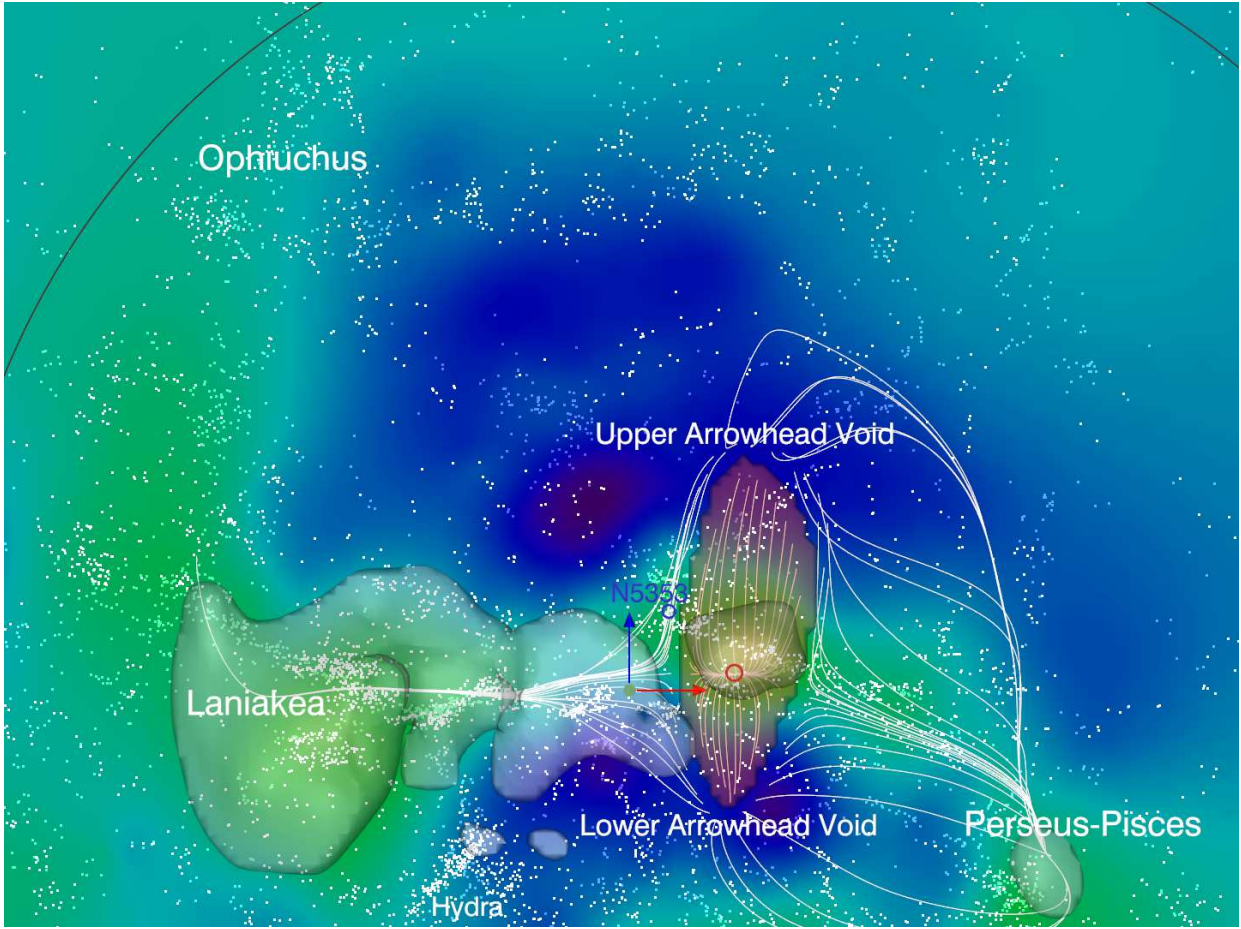}
\caption{Arrowhead Supercluster squeezed between Laniakea and Perseus-Pisces superclusters.  The view is of an $SGX-SGZ$ slice in the interval $1500<SGY<3500$~\kms. Shade of green and blue indicate levels of density, from higher to lower respectively.  Translucent contours show an isodensity surface.  Flow lines are shown for seeds near the inside and outside of the Arrowhead Supercluster, indicated by the dark surface.  White dots are galaxies within the slice drawn from the XSCz catalog.}
\label{Lani-PP}
\end{center}
\end{figure}

The flow patterns shown in Figure~\ref{sgxy} were obtained by centering at [1500, 2625, 500]~\kms\ in supergalactic coordinates and extracting the Wiener Filter density field within a radius of 10,000~km/s, with densities outside this sphere set to the mean density.\footnote{Flow patterns are constructed by evaluating the Wiener Filter velocity field at a seed position, then displacing in the direction of that motion, reevaluating the velocity field at the new position, and so on, to wherever one is lead.}   The next sequence of images provide alternate views.  Figure~\ref{Lani-PP} provides a nice view of the competition between the Laniakea and Perseus-Pisces superclusters.  Coma is to the background of the slice of an SGX-SGZ plane shown here.  The dark jagged surface outlines the limits of the Arrowhead Supercluster.  The translucent surfaces, including one inside the Arrowhead, illustrate an isodensity surface from the Wiener Filter reconstruction.  It is seen that there are voids above and below (plus and minus SGZ) the Arrowhead, here called Upper and Lower Arrowhead voids, which allows the domain of dominance of the Arrowhead to expand in those directions.

\begin{figure}[!]
\begin{center}
\includegraphics[scale=0.8]{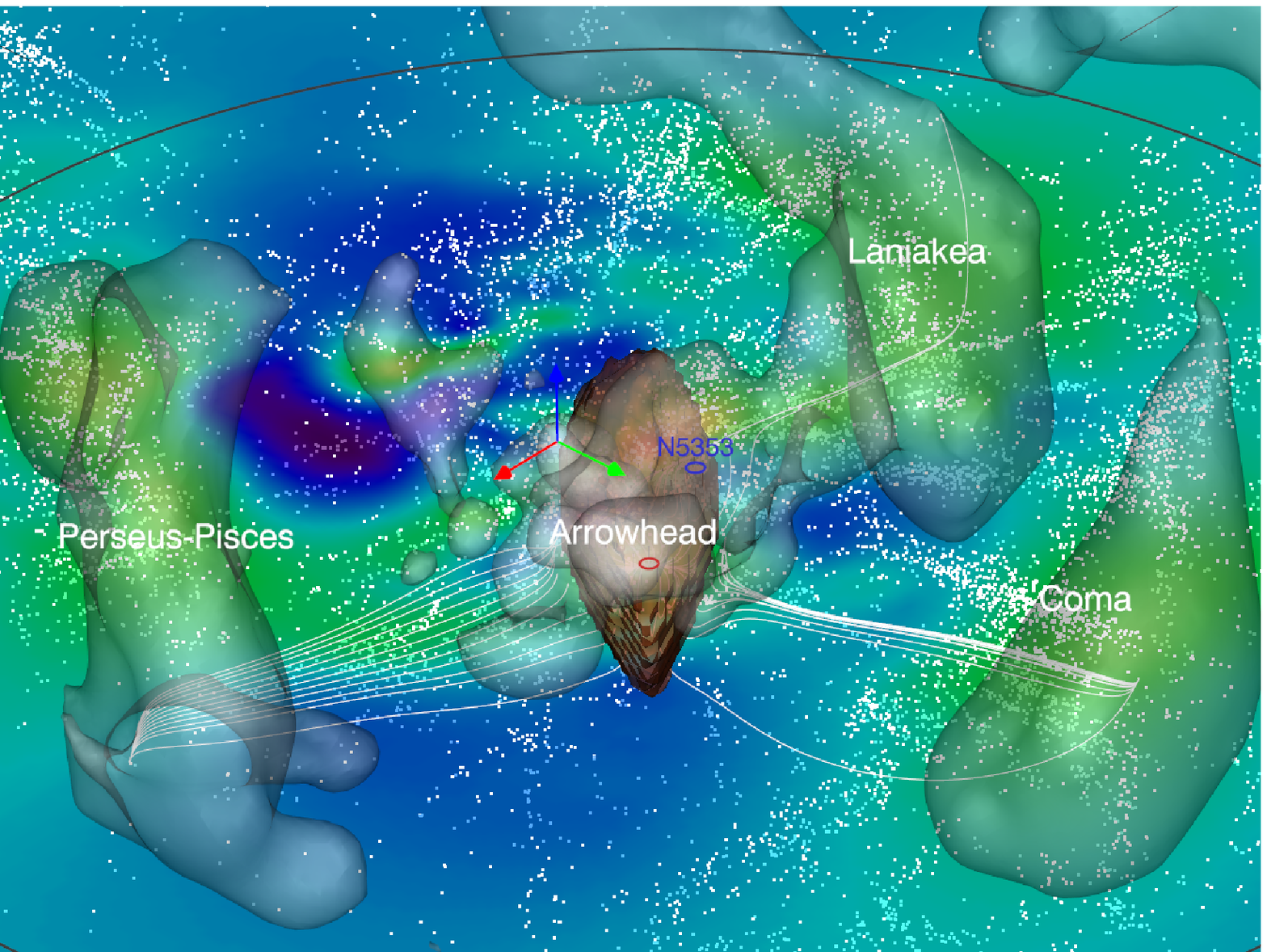}
\includegraphics[scale=0.8]{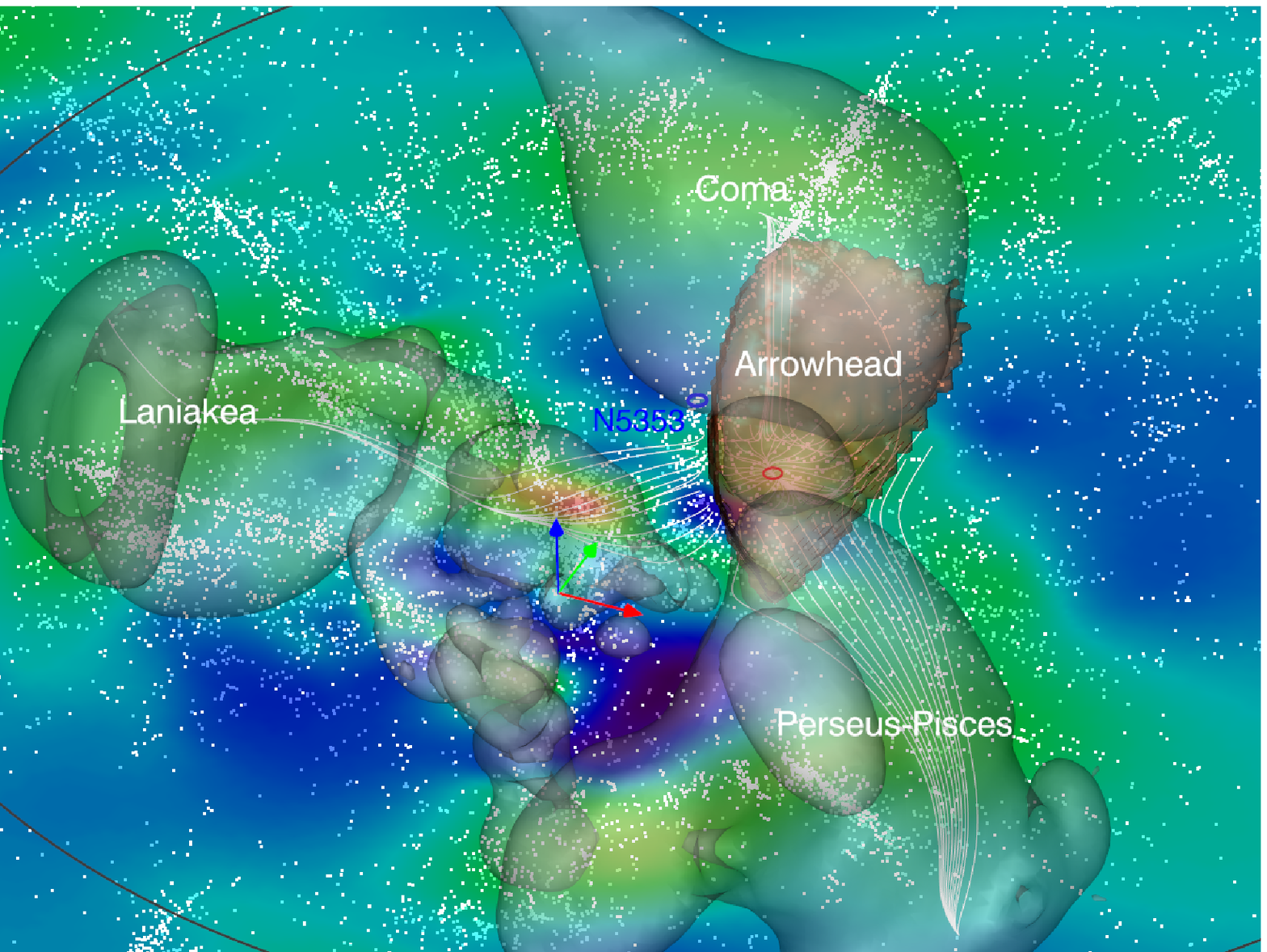}
\caption{Two views of flow patterns from seeds external to the Arrowhead Supercluster.  {\it Top:} competitive pulls between Perseus-Pisces and Coma structures, with Laniakea in the background. {\it Bottom:} in this rotated view, Laniakea's pull to the left is seen to be complemented by a push from a void at the right.}
\label{PP-Coma_ortho}
\end{center}
\end{figure}

The competition between the three external attractors is further explored in the pair of images in Figure~\ref{PP-Coma_ortho}.  The top panel shows the competing attractions of Perseus$-$Pisces and Coma.  Perseus$-$Pisces appears to have the greater influence.  The Wiener Filter reconstruction indicates there is a continuous over-density running to Perseus$-$Pisces although the continuity is obscured in the observed distribution of galaxies by the plane of the Milky Way.  On the other hand, the gap between the Arrowhead and Coma is under-dense, albeit laced with connecting filaments \citep{2008AJ....135.1488T}.  In the bottom panel it is seen that Laniakea gets help rather than hindrance from an opposed void we call "Far Arrowhead".  With close inspection it can be seen that there is also a void between Laniakea and the Arrowhead,  the "Virgo Void" in \citet{2013AJ....146...69C},\footnote{In the catalog of nearby voids by \citet{2013AstBu..68....1E} the Virgo Void is void 79, the closest part of the Far Arrowhead Void is their void 91, the Lower Arrowhead Void is their void 18, while the Upper Arrowhead Void is beyond the limit of their catalog.} though threaded by the CVn Spur. 

The robustness of the isolation and dimensions of the Arrowhead feature can be evaluated by varying the radius of the sphere governing the separation into local and tidal flows.  The local flow pattern and extent is essentially unchanged over the range of radii from 7000 to 11,000 \kms.

\onecolumn
\twocolumn

\onecolumn

\begin{figure}[!]
\begin{center}
\includegraphics[scale=0.75]{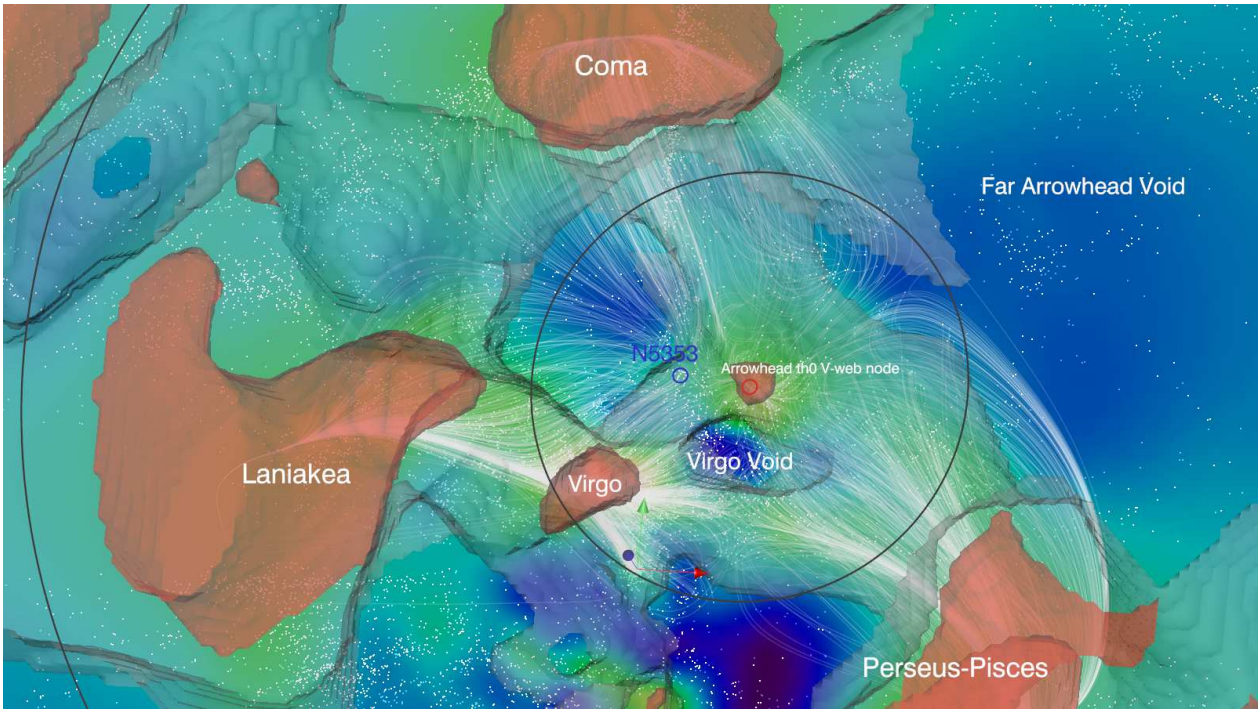}
\caption{The V-web.  The scene is a zoom of Fig.~\ref{sgxy} except here the translucent contours illustrate the surfaces enclosing knots (rust) and filaments (grey) as defined by eigenvalues of the shear tensor evaluated locally.  The red circle is at the center of the Arrowhead Supercluster.  Velocity flow lines in white are seeded at the locations of galaxies within the black ring with radius 3000~\kms.  The Milky Way is at the origin of the red arrow pointing right toward positive SGX, the green arrow directed up toward positive SGY, and the blue arrow pointing toward positive SGZ almost directly out of the page; each arrow 1000~\kms\ in length.}
\label{vweb_xy}
\end{center}
\end{figure}

\twocolumn

\subsection{The V-Web}

Figure~\ref{vweb_xy} is a reprise of the inner portion of Figure~\ref{sgxy} except that the translucent surfaces identify knots and filaments of the V-web rather than iso-density surfaces.  The idea of the V-web has been discussed by \citet{2012MNRAS.425.2049H}.  The V-web is defined by the shear tensor
\begin{equation}
\Sigma_{\alpha\beta}=-0.5 ({{\partial v_{\alpha}} \over {\partial r_{\beta}}} + {{\partial v_{\beta}} \over {\partial r_{\alpha}}})/{\rm H}_0  
\end{equation}
where $\alpha$ and $\beta$ are drawn from the three orthogonal directions.
The eigenvalues of the velocity shear tensor, $\lambda_{1,2,3}$ are evaluated locally on a grid, ordered from most positive to most negative.   If all three eigenvalues are positive the grid position is within a bound knot, if two are positive 
then the position is within a filament, one positive then it is part of a sheet, or if all eigenvalues are negative the position is in a void.  

The concept of the V-web provides an objective way to define a supercluster.  The center of a basin of attraction, the point of strongest convergence of the flow, lies at the grid point where $\lambda_3$ is greater than zero and has a local maximum.  This location may have a motion due to the influences of adjacent attractors.  For the Arrowhead, the location and velocity vector components at the bottom of the basin of attraction are given in Table~1.  Velocity uncertainties on each axis, evaluated from constrained realizations as discussed immediately below, are $\pm 60$ \kms.
The motion, mainly toward $-$SGX and $-$SGZ is in the general direction of our motion with respect to the cosmic microwave background frame.  There is a bulk motion of the entire region of $\sim 200$ Mpc around us \citep{2015MNRAS.449.4494H} and the Arrowhead participates in this large scale flow.

\begin{table}
\begin{center}
\caption{Arrowhead Central V-web Parameters}
\begin{tabular}{lrrr}
\hline
 In km s$^{-1}$ & SGX & SGY & SGZ \\
\hline
Pos'n, basin of attraction  &  1500 & 2625 & 500  \\
Velocity  of attractor                 &  -248 &     22 & -244 \\
\hline
\end{tabular}
\end{center}
\label{boa}
\end{table}

Comparison between 20 constrained realizations of Gaussian random fields \citep{1991ApJ...380L...5H,2014Natur.513...71T,2015MNRAS.449.4494H} can be used to evaluate the significance of the Arrowhead.  
The Wiener Filter is a Bayesian estimator of the underlying velocity field, given the data, the errors model and an assumed prior model. Here we assume the standard $\Lambda$CDM model. The constrained realizations samples the fluctuations (or residual) around the mean, i.e. Wiener Filter, field. The variance of the constrained realizations around the mean field provides a measure for the robustness of the Wiener Filter reconstruction.
Nearby, and including the region of the Arrowhead, observations highly constrain the fluctuations with the consequence that the separate realizations have similar characteristics.  At larger radii where observations are sparse the separate realizations can look quite different from each other.  The information provided by the run of constrained realizations is encoded in the dispersion at each cell.  Off the Galactic plane, out to roughly 10,000 \kms\ density fluctuations are constrained to better than the mean value of the standard deviation.
 
In a region within 1000 \kms\ of the location specified in Table~1, the position and amplitude of $\lambda_3$ is identified in each of the individual constrained realizations.  The value of $\lambda_3$ is positive in 19 of 20 cases, the characteristic of a knot.  The 20th case has the characteristics of a filament within the immediate region of the mean basin of attraction.  The average deviation in position of the basin of attraction between constrained realizations is 717 \kms.

\begin{figure}[t!]
\begin{center}
\includegraphics[scale=0.37]{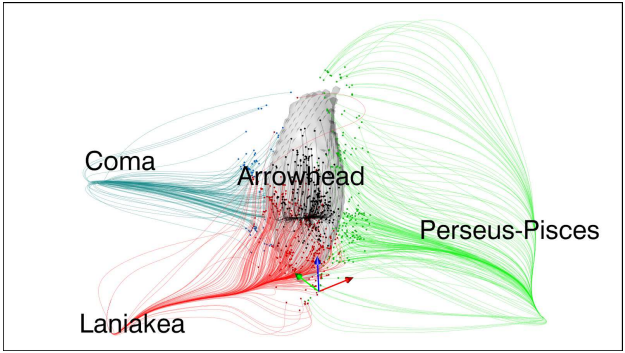}
\caption{Flows away from the edges of the Arrowhead Supercluster toward alternatively the Laniakea, Perseus-Pisces, and Coma basins of attraction.  The Arrowhead is bounded by voids on the far side from Laniakea and in the directions above and below in SGZ. The orientation of the plot is given by the 3 arrows located at the position of the Milky Way with lengths of 1000~\kms: red arrow toward +SGZ, green arrow toward +SGY, and blue arrow toward +SGZ.}
\label{extern_flows}
\end{center}
\end{figure}

The dominant characteristic of the Arrowhead Supercluster is that of a filament, aligned with the flow pattern internal to the structure.  The most prominent structure seen in the redshift space distribution of galaxies, the CVn$-$Cam Cloud, runs parallel to the V-web structure but offset to the near edge due to redshift distortion.  The seeded flows external to the Arrowhead are again seen in Figure~\ref{vweb_xy} to run toward either Perseus$-$Pisces at 4 o'clock, Coma at 11 o'clock, or Laniakea at 9 o'clock, with the prominent Far Arrowhead Void at 2 o'clock.  There is the modest Virgo Void between Arrowhead and Laniakea at 7 o'clock and an under dense region between Arrowhead and Coma.  The bridge toward Perseus$-$Pisces seems more robust, although tucked behind galactic obscuration.  As was seen in Figure~\ref{Lani-PP}, there are the Upper and Lower Arrowhead voids above and below the Arrowhead in SGZ.  The competition at the boundary of the Arrowhead is captured in the simplified Figure~\ref{extern_flows} where flows from seeds beyond the extremity of the Arrowhead are followed to their sources.  Situated in the Arrowhead Supercluster, an observer would see attractors in three cardinal directions and voids in the remaining three cardinal directions.

Various of the figures identify the location of the NGC~5353/54 Group (shortened to the notation N5353).  This group with a mass of $3 \times 10^{13}~\Msun$ at 35~Mpc (redshift distance 2645~\kms) is the dominant knot in the Arrowhead region \citep{2008AJ....135.1488T}.  It is, however, at the very edge of the Arrowhead basin of attraction as defined by velocity flows.  The NGC 5353/54 Group lies at the boundary with the Laniakea pattern, near the juncture with Coma flows.  Indeed, a large fraction of the galaxies associated with the Arrowhead Supercluster are at the near boundary, just beyond the Virgo Void and aligned with the axis from Perseus-Pisces to Coma.  

\subsection{A Closer Look}

In order to evaluate the internal structure of the Arrowhead Supercluster it is advantageous to convert to a coordinate system compelled by the local distribution of galaxies.  The V-web is a guide. The velocity shear eigenvalues in the region define a filament within a sheet.  The long axis of the filament specifies the principal axis in the Arrowhead frame "AX", taken to be positive in the direction toward Perseus$-$Pisces and negative toward Coma.  The minor axis in the Arrowhead frame "AZ" is taken to be normal to the sheet, positive (negative) in the direction toward the Upper (Lower) Arrowhead Void.  The intermediate axis "AY" lies normal to the filament and in the sheet.  Laniakea lies roughly in the $-$AY direction with the Far Arrowhead Void toward +AY.

\begin{figure}[!]
\begin{center}
\includegraphics[scale=0.25]{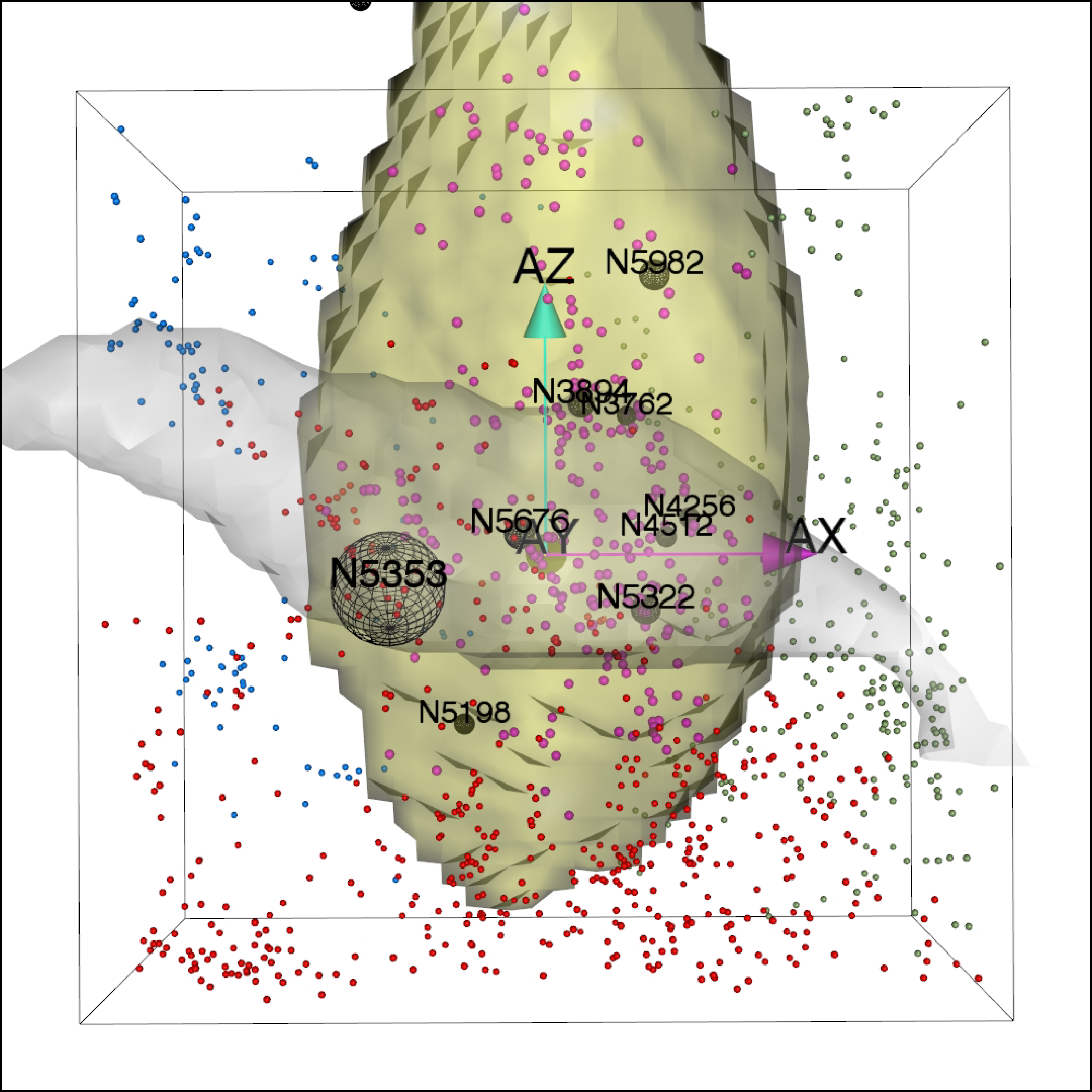}
\includegraphics[scale=0.25]{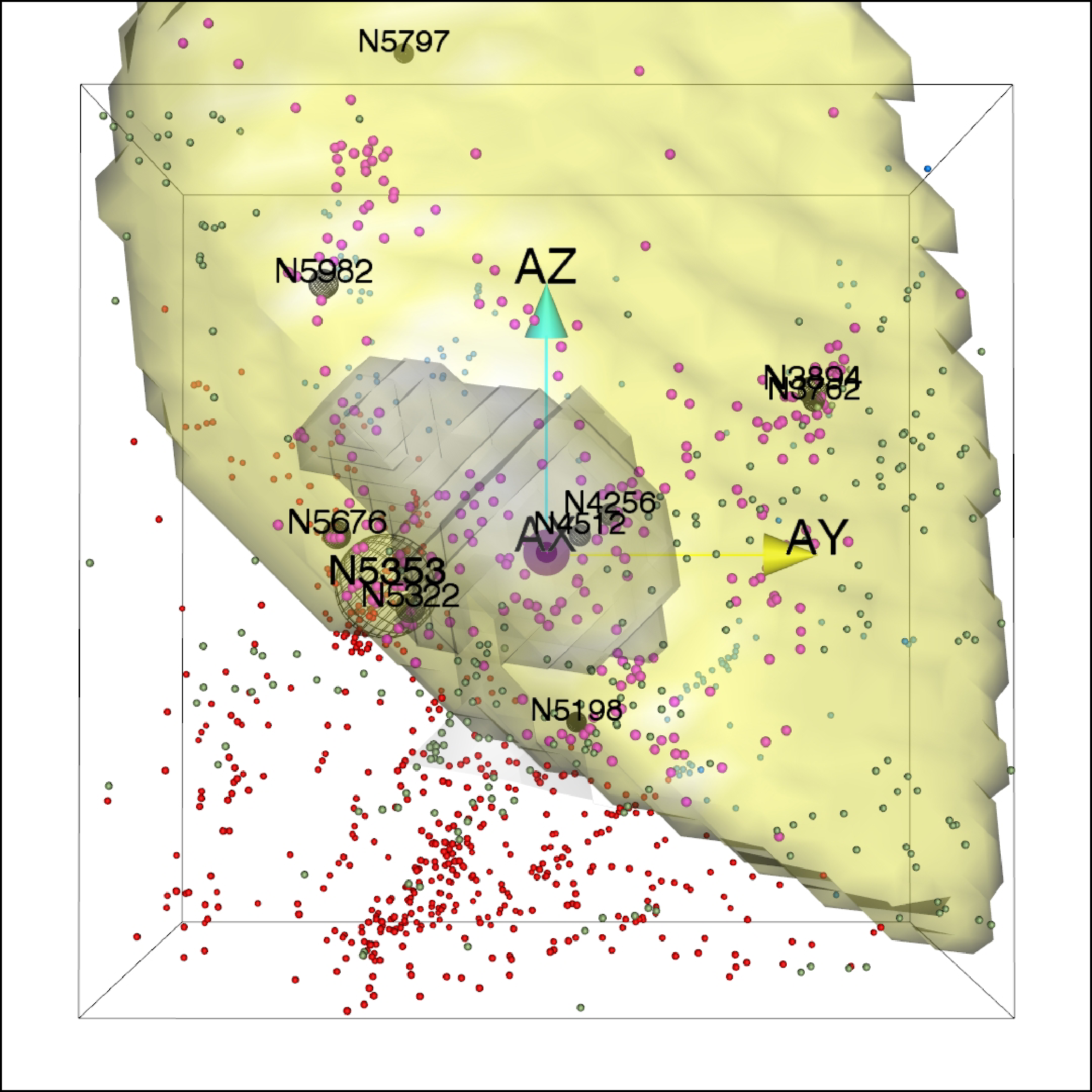}
\caption{Two orthogonal views of the Arrowhead Supercluster edge-on to the plane of the structure as prescribed by the V-web analysis.  The views are normal to the filament in the top panel and along the filament in the bottom panel (bounded by the grey surface in each panel).  The yellow surface is the boundary of the Arrowhead defined by local flow patterns.  Individual galaxies within the Arrowhead are identified by magenta points.  Galaxies outside Arrowhead are in red if they participate in flows toward Laniakea, green if toward Perseus$-$Pisces, and blue if toward Coma.  The location of the NGC 5353/54 Group and 8 smaller groups are identified by the gridded spheres, scaled roughly to virial radii.  The cardinal axes in the Arrowhead frame AX, AY, AZ are colored magenta, yellow, and cyan respectively.  The lengths of the vectors are 1000~\kms.}
\label{ax-ay}
\end{center}
\end{figure}

Two views in the natural coordinate system of the Arrowhead Supercluster are seen in Figure~\ref{ax-ay}.  The top panel is a view edge-on to the V-web sheet, oriented to capture the long axis of the V-web filament, running horizontally to the right from the vicinity of the NGC 5353/54 Group.  The view in the bottom panel is also edge-on to the sheet, now looking down the axis of the filament where several groups pile up.  Individual galaxies are identified; those within the Arrowhead in magenta and those on the periphery in other colors, coded by their flow affiliation.

To gain an understanding of the structure inside and adjacent the Arrowhead Supercluster it is particularly useful to give attention to the accompanying video.  The structure is lumpy.  There are only stunted filaments.  A substantial fraction of the galaxies lie at the near boundary.  The NGC 5353/54 Group is a case in point.  This, the largest collapsed structure in the vicinity with mass $3 \times 10^{13}~\Msun$ \citep{2008AJ....135.1488T} nominally spills across the Arrowhead boundary on the side captured by the Laniakea Supercluster.   Stepping down a factor three in mass (Tully 2015), there are two groups comfortably within the Arrowhead (brightest galaxies NGC 3894 and NGC 5982) and two groups (NGC 5322 and NGC 5676) on the front surface close to NGC 5353/54.  With another factor 3 step down in mass, four additional groups are found within the Arrowhead (NGC 3762, NGC 4256, NGC 4512, NGC 5198) and one more on the edge (NGC 5797), on the side toward Coma.

\onecolumn
\twocolumn

It has been stated that the galaxies in the Arrowhead accumulate toward the near boundary and stretch along the axis between the Coma and Perseus$-$Pisces structures.  Yet the boundaries of the Arrowhead are most truncated along the AX axis, in the line between Coma and Perseus$-$Pisces, and in the orthogonal direction toward Laniakea.  The Arrowhead is most truncated toward gravitational competitors.  The boundaries of the Arrowhead extend in the directions toward the voids where there are few galaxies but no competition.

\subsection{Bridges to Adjacent Structures}

\begin{figure}[!]
\begin{center}
\includegraphics[scale=0.41]{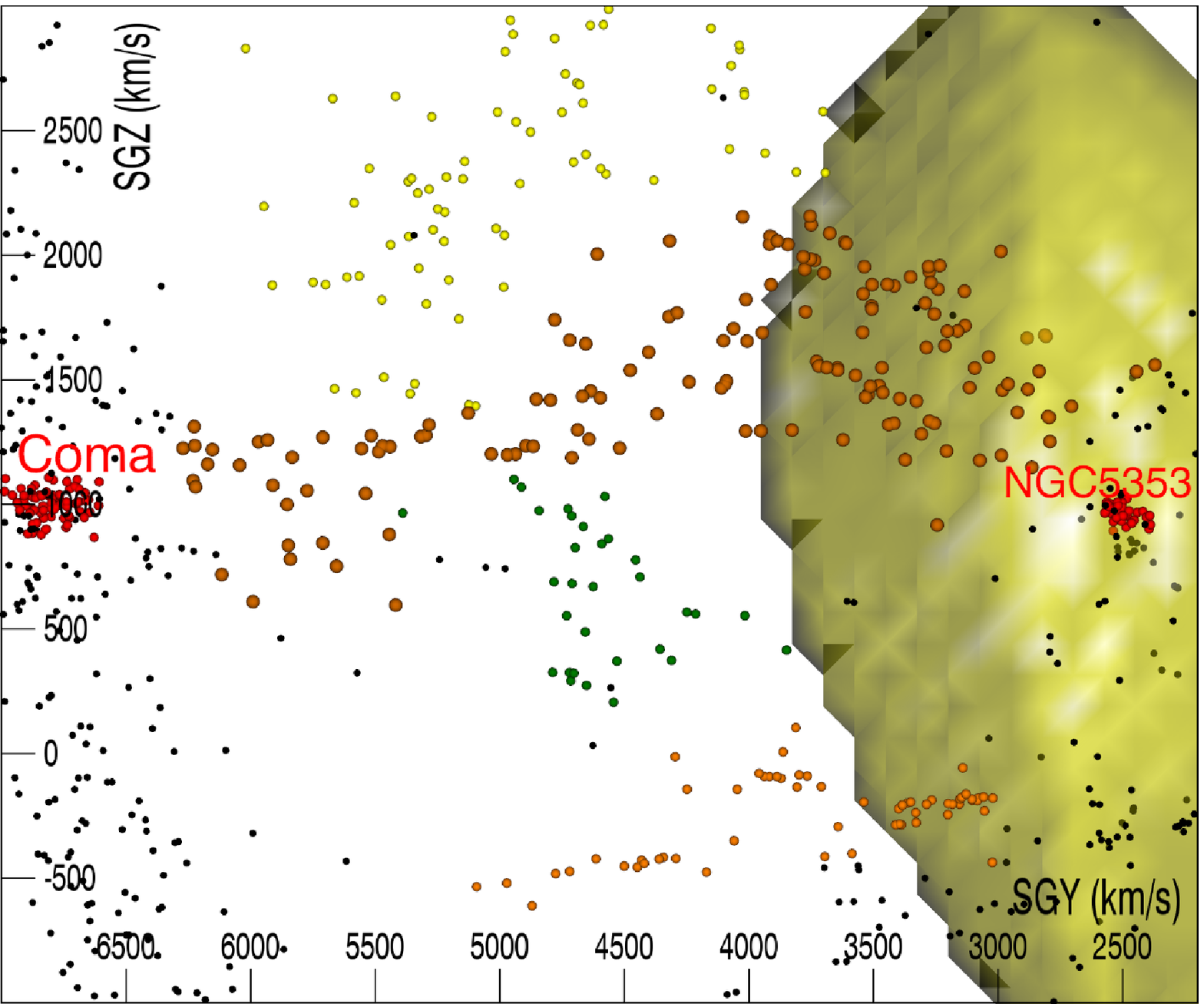}
\includegraphics[scale=0.575]{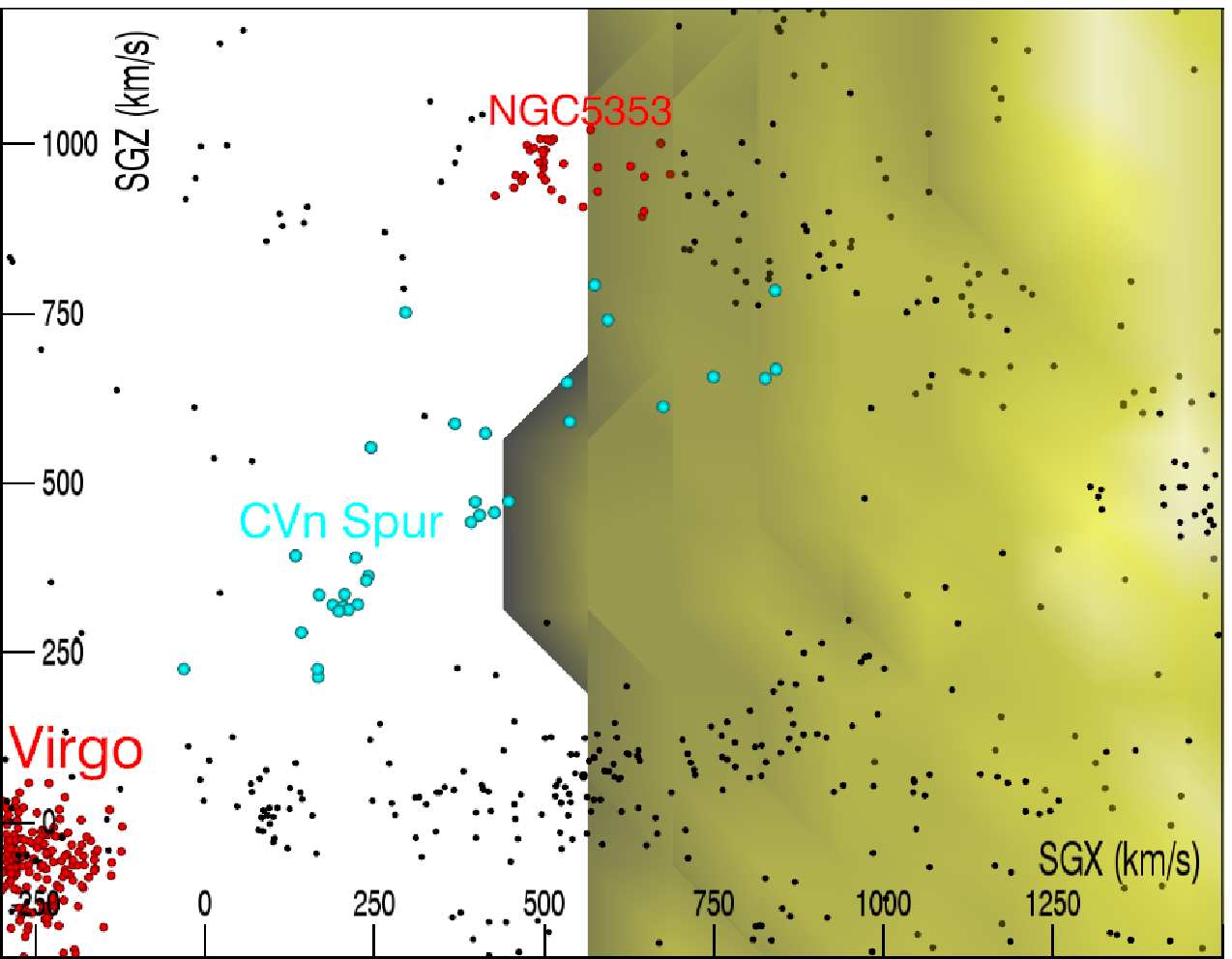}
\includegraphics[scale=0.615]{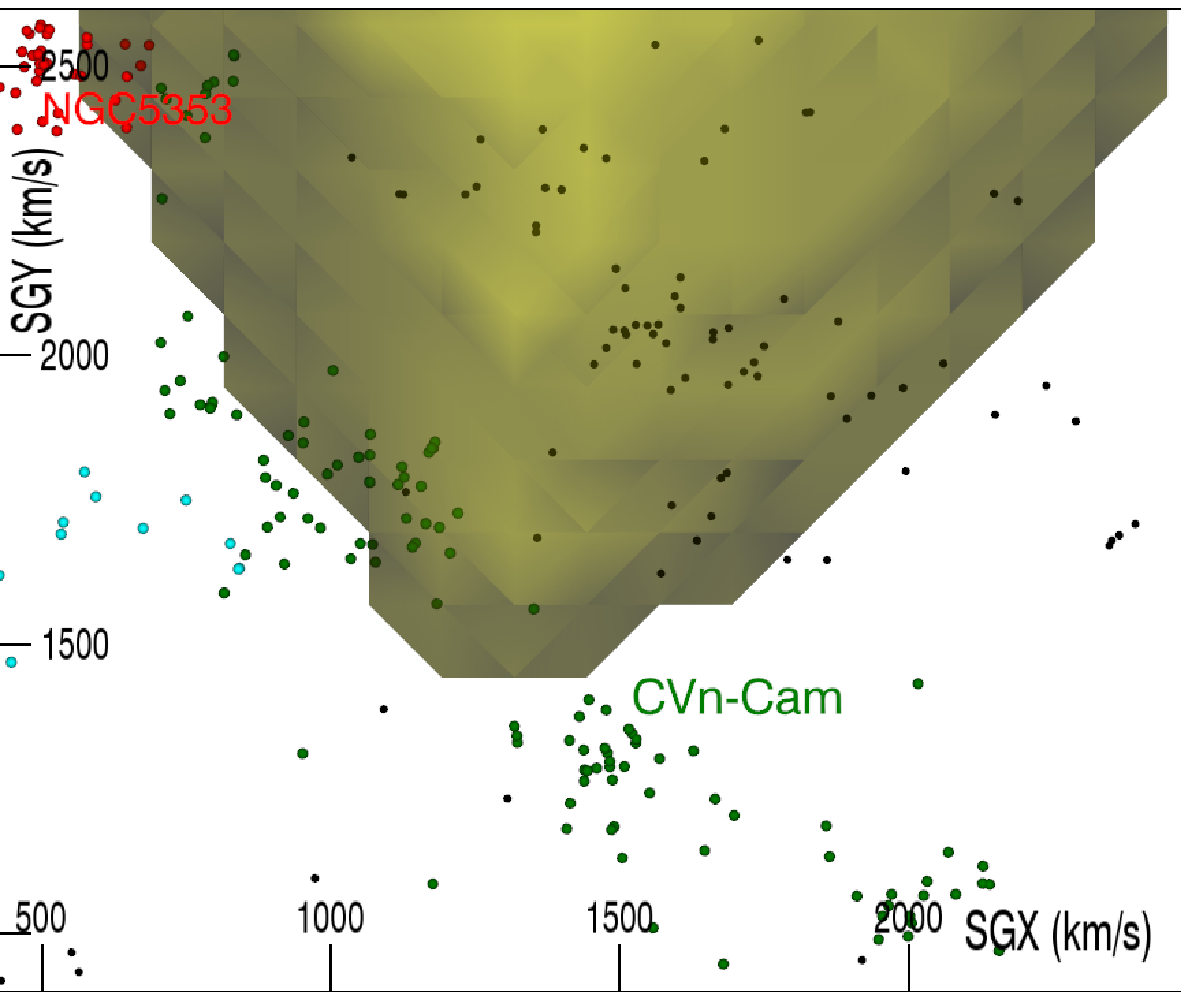}
\caption{Filaments between the Arrowhead Supercluster and adjacent structures. {\it Top.} Filament in orange linking the region of the NGC 5353/54 Group and the Coma Cluster.  {\it Middle.}  CVn Spur, a filament in cyan between the Arrowhead and the Virgo Cluster.  {\it Bottom.}  The CVn$-$Cam Cloud in green that runs from near the NGC 5353/54 Group downward to the right along the edge of Arrowhead toward the Perseus$-$Pisces structure.}
\label{bridges}
\end{center}
\end{figure}

The filamentary links between the Arrowhead Supercluster and the bordering structures are shown in greater detail in Figure~\ref{bridges}.  The top two panels are variants of Fig.~2 in the paper by \citet{2008AJ....135.1488T} discussing the NGC 5353/54 Group.  The top panel gives emphasis to a filament that connects the NGC 5353/54 Group and the Coma Cluster.  The middle panel gives focus to a filament running from near the NGC 5353/54 Group to the Virgo Cluster, a structure called the Canes Venatici Spur in the {\it Nearby Galaxies Catalog} \citep{1988ngc..book.....T}.  Finally, in the bottom panel the emphasis is on what was called the Canes Venatici$-$Camelopardalis Cloud in the {\it Nearby Galaxies Catalog}.  This important filament starts near the NGC 5353/54 Group, runs along the near edge of the Arrowhead Supercluster, then continues toward the Perseus$-$Pisces region.

\section{Summary}

It has been argued \citep{2014Natur.513...71T} that shears in the velocity field of galaxies provides a quantitative way to define superclusters. The recipe is (1) calculate the velocity shear tensor, (2) evaluate the V-web and identify the attractors, (3) locate the bottom of a basin of attraction where the smallest eigenvalue of the shear tensor is maximal, then (4) identify the full domain of the basin of attraction by construction of the flow field in the frame of reference of the bottom of the basin of attraction.  Operationally, the velocity field inferred from the Wiener Filter density reconstruction can be separated into local (divergent) and tidal components \citep{2012ApJ...744...43C, 2014Natur.513...71T}.
 Local basins of attraction can be identified that metaphorically can be related to terrestrial watersheds.  Boundaries to local flow patterns can be identified.  A volume within a bounded region can be identified with a single supercluster.  

Accordingly, we live in the Laniakea Supercluster.  Our Galaxy is near an edge of Laniakea, near boundaries with structures in Coma and Perseus$-$Pisces.\footnote{We are not formally referring to Coma and Perseus$-$Pisces as `superclusters' because present distance information is insufficient to delimit these entities on their back sides.  Each of these structures may be part of something larger.}   Velocity field information reveals that there is a region between these three major attractors that stands apart.  There are flows from its boundary toward each of Laniakea, Coma, and Perseus$-$Pisces but within the boundary local flows are captured.  The region is relatively nearby and at high Galactic latitude so it is amenable to investigation.  The dimensions of the region are roughly $1800 \times 3000 \times 3800$ \kms\ and the contained mass is roughly $10^{15}~\Msun$, linear dimensions a factor of 5 less than Laniakea and a mass two orders of magnitude less.  We call this entity the Arrowhead Supercluster. 

The dimensions of the Arrowhead are reflective of its environment.  The short dimension within the Arrowhead basin of attraction is the principal axis of the V-web filament defined by local values of the shear tensor, the axial direction called AX.  This axis is aligned with the line between Coma and Perseus$-$Pisces and is the shortest dimension because of competition with the two larger attractors.  The intermediate dimension, AY, is within a sheet defined by the V-web eigenvalues.  It is directed on the near side toward Laniakea while what is found on the far side is a void, here called the Far Arrowhead Void.     The Arrowhead is most extended in the third orthogonal direction, AZ.  The limits along this axis are in voids, called the Upper and Lower Arrowhead voids.  Dimensions are truncated in the directions of greatest gravitational competition and largest in directions toward voids.

The distribution of galaxies in the vicinity of the Arrowhead Supercluster agrees with the V-web picture.  Galaxies tend to pile up on the near edge, toward Laniakea and away from the Far Arrowhead Void, a real effect because the XSCz redshift catalog used for comparison has a high level of completion in this distance range.  There is a major concentration of galaxies toward the Coma to Perseus$-$Pisces alignment.  The real estate toward the three voids is relatively empty.   The most important collapsed region, the NGC 5353/54 Group with $3 \times 10^{13}~\Msun$, is at the edge of the Arrowhead Supercluster boundary.  Likewise, the CVn$-$Cam Cloud seen in the lower panel of Figure~\ref{bridges} skirts the Arrowhead boundary before clearly separating toward Perseus$-$Pisces.  Structure within the Arrowhead is clumpy but ill defined.  A group catalog \citep{2015AJ....149..171T} of galaxies from the 2MASS redshift survey \citep{2012ApJS..199...26H} identifies 9 groups in the region with 20\% to 50\% of the 2MASS galaxy content of the NGC 5353/54 Group.   There is no noteworthy concentration of these groups.  The Arrowhead has no evident core.

The Arrowhead Supercluster is tiny compared to its neighbors.  It is suspected that it owes its continued existence to a standoff in the gravitational competition between three big neighbors: Laniakea, Coma, and Perseus$-$Pisces.  The Wiener Filter density field reconstruction suggests that there is a significant filament running from the Arrowhead toward Perseus$-$Pisces, albeit through the zone of obscuration.  Filaments also run from the Arrowhead toward Coma and Laniakea.   The placement of voids is relevant.  An observer in the Arrowhead Supercluster looking in 6 orthogonal directions sees Coma and Perseus$-$Pisces opposed to each other along the principal axis of structure, a filament defined by the velocity shear eigenvalues, then Laniakea in an orthogonal alignment within a sheet defined by the velocity shear and opposed by the large Far Arrowhead Void, and finally in the remaining two directions the observer sees the Upper and Lower Arrowhead voids.  Even in the directions toward attractors voids are important.  The region between the Arrowhead and Coma is quite under dense.  And although the Arrowhead Supercluster is close to Laniakea in the vicinity of the Virgo Cluster it is distinctly separated by the Virgo Void. 

\smallskip\noindent
Accompanying video: http://irfu.cea.fr/arrowhead

\bigskip
\noindent
{\bf Acknowledgements}

Financial support for the Cosmicflows program has been provided by the US National Science Foundation award AST09-08846, an award from the Jet Propulsion Lab for observations with Spitzer Space Telescope, and NASA award NNX12AE70G for analysis of data from the Wide-field Infrared Survey Explorer.  Additional support has been provided by the Israel Science Foundation (1013/12) and the Lyon Institute of Origins under grant ANR-10-LABX-66 and the CNRS under PICS-06233.
  
\bibliography{paper}
\bibliographystyle{apj}

\
\end{document}